\documentclass[a4paper,fleqn]{cas-dc}

\usepackage[numbers]{natbib}
\usepackage{graphicx}
\usepackage{adjustbox}
\usepackage{amsmath,amssymb,amsfonts}
\usepackage{booktabs}
\usepackage{capt-of}

\begin{document}
\let\WriteBookmarks\relax
\def\floatpagepagefraction{1}
\def\textpagefraction{.001}

\shorttitle{MARS}    
\shortauthors{H. Choi}  

\title [mode = title]{MARS: Malware Analysis with Rule-Based Scoring of LLM Claims}  

\author{Hyeongjun Choi}[orcid=0000-0003-1700-3729]
\ead{hjchoi@78researchlab.com}
\affiliation{organization={78ResearchLab},
            addressline={8, Yangpyeong-ro 25-gil, Yeongdeungpo-gu},
            city={Seoul},
            postcode={07207},
            country={Republic of Korea}}

\begin{abstract}
Large language models can triage malware through direct verdicts or behavioral claims scored by an external policy. We present MARS, a malware triage framework, and compare direct classification with single-pass claim scoring using the same evidence collector and identical static evidence bundles for each model. The evaluation covers 1,195 PE and ELF binaries grouped into 1,001 near-duplicate clusters and six language models, with deterministic rules providing a baseline. Direct classification is more accurate for all six models. On samples with usable outputs from both paths, its accuracy advantage ranges from 3.7 to 20.9 percentage points, with all 95\% cluster-bootstrap confidence intervals for the differences above zero. It also achieves higher malicious alert recall in ten of twelve platform and model combinations. Claim mediation provides no consistent reduction in performance variation across models. Separate subset studies find more consistent alert decisions for direct classification and a larger recall loss for the claim path when predefined indicator fields are removed. In a family identification probe, claims yield higher accuracy than verdict labels but lower accuracy than evidence text. Retained claims expose the inputs to verdict computation and permit policy revision without another model call. We reproduce archived verdicts exactly and apply a revised policy to the same records, including outputs from two additional models withdrawn by their provider. Under the evaluated claim taxonomy and additive policy, these results favor direct classification when only a verdict is required, while demonstrating that retained claims support explicit policy inspection and revision.
\end{abstract}


\maketitle

\section{Introduction}
Malware triage requires a decision about an unfamiliar binary from evidence that is often incomplete and difficult to reconcile. Analysts combine file metadata, imports, strings, code, and runtime observations to assess whether a program is malicious~\cite{aonzo2023humans}. Packing can conceal relevant static content, while dynamic analysis observes only behavior triggered in a particular environment~\cite{muralidharan2022file,afianian2019malware}. The resulting verdict depends on how these observations are interpreted and combined.

Large language models (LLMs) have been applied to malware analysis and reverse engineering to interpret code and analysis artifacts~\cite{xu2024large,zhang2025llms}. A direct classifier asks the model to integrate the collected evidence and return a label, optionally with a rationale. This approach leaves evidence interpretation and verdict assignment to the same model response. The plausibility of the accompanying rationale does not establish its faithfulness to the decision process~\cite{jacovi2020faithfulness}.

An alternative separates interpretation from verdict assignment. The LLM produces typed behavioral claims, and a fixed downstream policy scores those claims to determine the label. The policy consumes the same records retained for inspection, making each claim's contribution to the score explicit. This use of an intermediate representation is related to concept bottleneck models, which predict concepts and use them to determine the final label~\cite{koh2020concept}. In the setting studied here, an LLM generates behavioral claims without concept-level supervision, and a prespecified scoring policy assigns the verdict. Stored claims can then be rescored without another model call. This interface does not establish the semantic correctness of the claims or eliminate variability in their generation.

Comparisons between independently developed malware analysis systems are complicated by differences in evidence collectors, models, and inputs, making observed differences difficult to attribute to the decision architecture. This paper evaluates the trade-offs between direct LLM classification and classification through typed intermediate claims scored by a fixed downstream policy while holding the evidence collector and LLM constant. The evaluation considers classification performance, stability, and the information and rescoring capabilities provided by the retained outputs.

We propose MARS, a malware triage framework for Portable Executable (PE) and Executable and Linkable Format (ELF) binaries that separates LLM-based evidence interpretation from verdict assignment by an explicit scoring policy. MARS serializes collected artifacts into a common evidence bundle and implements three decision paths for controlled comparison, consisting of deterministic rules, direct LLM classification, and structured claim extraction followed by scoring. All paths use the labels benign, suspicious, malicious, and unknown. The main experiment compares these paths on 1,195 binaries using basic static evidence, with the two LLM paths evaluated across six models. The claim path uses a confidence-free additive score and a benign threshold calibrated separately for each model on a disjoint development set. Direct classification uses the model's label without a downstream threshold. The comparison concerns the complete decision paths, including their prompts and output schemas.

Direct classification achieves higher malicious alert recall in ten of the twelve platform and model combinations and higher overall accuracy on all six models. Cluster-bootstrap 95\% confidence intervals for the paired accuracy differences exclude zero for every model. Claim mediation also exhibits greater variation across models in benign handling and malicious alert recall. These findings apply to the evaluated evidence collector, claim taxonomy, and scoring policy.

The evaluation also examines what each path retains beyond its verdict. In an auxiliary family-identification probe, claim sets outperform verdict-only representations but score below the raw evidence text, while direct justifications achieve scores close to those of the claim sets. Repeated execution shows that a fixed scoring policy does not ensure end-to-end stability. Stored claims nevertheless permit exact replay of the policy calculation and rescoring under alternative policies, including after the generating model becomes unavailable. A prespecified perturbation study finds greater recall loss from indicator hiding on the claim path than on the direct path. A separate follow-up evaluates runtime evidence on selected cases using static-plus-decompiler evidence as the control. The claims are evaluated as inputs to the verdict policy, without expert validation of their semantic accuracy or measurement of their benefit to analysts.

Our contributions are as follows.

\begin{itemize}
    \item We present MARS and a matched comparison of direct classification and claim-mediated scoring for PE and ELF binaries across six LLMs, with a common evidence collector, identical evidence bundles, and a deterministic rules-only baseline.
    \item We report an accuracy advantage for direct classification over the evaluated claim path, supported by cluster-based uncertainty estimates and analyses of architecture strata, metadata difficulty, and threshold selection.
    \item We characterize the trade-offs of the claim interface through processing cost, family-information probes, repeated execution, policy replay, and evidence perturbation, distinguishing the reproducibility of a stored policy calculation from the stability of the complete pipeline.
\end{itemize}

\section{Background}
\subsection{Bounded Observation Context}
Automated malware analysis provides partial observations of program behavior. Static artifacts may expose capabilities without showing whether they are exercised, while monitored execution records only behavior triggered under the supplied conditions. Packing and analysis evasion can further restrict what the collector observes~\cite{muralidharan2022file,afianian2019malware}. The resulting evidence bundle is a bounded observation context. A sample-level benign or malicious label supports evaluation of the verdict but does not establish the completeness of the evidence or the correctness of every inferred behavior.

Studies of analyst practice document the use of multiple tools and evidence sources during an investigation~\cite{yong2021inside,botacin2024malware}. Human analysts and automated classifiers may also rely on different feature groups when examining the same samples~\cite{aonzo2023humans}. A controlled comparison of downstream decision strategies must distinguish their effects from differences in the collected evidence. Holding the evidence bundle fixed permits this comparison, although any claims generated from it remain interpretations of the observations available to the model.

\subsection{Structured Outputs and Faithfulness}
An explanation is faithful to the extent that it reflects the process responsible for a prediction~\cite{jacovi2020faithfulness}. Local attribution methods such as LEMNA approximate a security classifier's behavior around an input to identify influential features~\cite{guo2018lemna}. Chain-of-thought prompting generates intermediate natural language reasoning before a final answer~\cite{wei2022chain}. The presence of this text does not establish faithfulness. Turpin et al.\ showed that explanations can omit factors that influence the answer~\cite{turpin2023unfaithful}. By intervening on reasoning text, Lanham et al.\ found that the extent to which models rely on it varies across tasks and models~\cite{lanham2023measuring}.

Claim-mediated scoring makes the role of the intermediate output explicit. A policy outside the LLM computes the score from specified claim fields. In the evaluated MARS claim path, these are the category and severity fields. Retaining the claims therefore permits inspection and replay of the score calculation. This procedural linkage does not establish semantic correctness or reveal the model's internal reasoning. It also guarantees reproducibility only for a fixed claim set and policy, since claim generation and severity assignment can vary across model calls.

\subsection{Claim Decomposition and Support}
Atomic claim decomposition allows supported and unsupported statements within a response to be assessed separately. FActScore measures the proportion of individual facts supported by a reference knowledge source~\cite{min2023factscore}. SAFE uses search results to assess support for individual facts~\cite{wei2024long}, while FacTool applies external tools to factuality checks across several domains~\cite{chern2023factool}. These approaches make support assessment conditional on the reference information and tools available to the evaluator.

Self-Refine and Chain-of-Verification show that iterative feedback or verification steps can improve generated responses on the tasks studied~\cite{madaan2023selfrefine,dhuliawala2024chain}. Their results motivate evaluating additional model calls, but do not make a subsequent call an independent source of ground truth.

The optional support assessment in MARS receives the candidate claims and the same evidence bundle used for extraction. It assigns the states supported, hypothesis, and unsupported to indicate direct, incomplete, and absent support in the supplied artifacts, respectively. These states are judgments made by the model. They do not establish that a behavioral claim is correct across possible executions or that unobserved behavior is absent. Because extraction and assessment use the same model family and overlapping context, their errors may be correlated. Under the evaluated policy, these support states are retained as annotations and do not affect verdict assignment.

\section{Methodology}
\subsection{System Overview}
MARS processes PE and ELF binaries through a shared evidence collector and alternative decision paths. Basic static artifacts, optionally augmented with decompiler output and runtime observations, are normalized into an evidence bundle. The rules-only path assigns a label through deterministic predicates. The direct path requests a label from an LLM. The single-pass claim path extracts typed claims for a fixed scoring policy. An optional second LLM pass attaches support annotations to the extracted claims.

\begin{figure*}[pos=t]
\centering
\includegraphics[width=\textwidth]{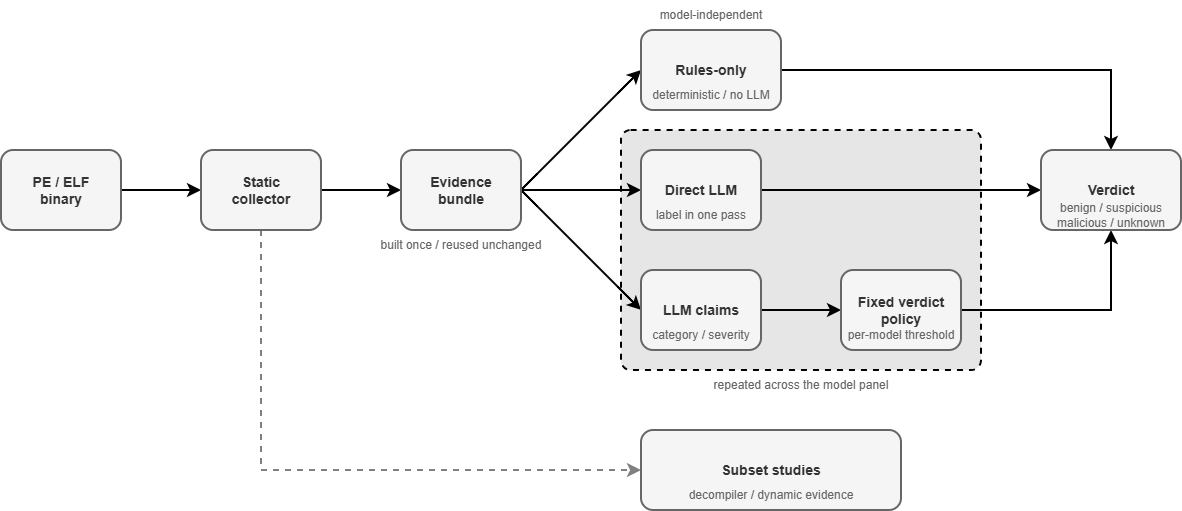}
\caption{MARS evaluation architecture. The full-corpus comparison uses a shared static evidence bundle for rules-only classification, direct LLM classification, and single-pass claim scoring. The two LLM paths are evaluated across six models. Additional evidence and support assessment are examined in subset studies.}
\label{fig:mars_overview}
\end{figure*}

The full-corpus comparison evaluates the rules-only, direct, and single-pass claim paths with basic static evidence. The rules-only baseline is run once, and the two LLM paths are evaluated across six models. The repeated-run study examines direct classification and two independently executed claim configurations across three models. A separate follow-up with Gemini 2.5 Flash examines runtime evidence on selected cases, using static-plus-decompiler evidence as the control. Figure~\ref{fig:mars_overview} and Table~\ref{tab:processing_configurations} summarize these configurations. Narrative reports, threat-type classification, ATT\&CK mapping, and response recommendations are outside the evaluation scope.

\begin{table*}[t]
\centering
\caption{Evaluated processing configurations. All configurations include basic static evidence. The dynamic follow-up uses static-plus-decompiler evidence as the control on 20 selected cases.}
\label{tab:processing_configurations}
\small
\adjustbox{max width=\textwidth}{%
\begin{tabular}{lllll}
\toprule
Path & Additional evidence & Support pass & Verdict assignment & Evaluation scope \\
\midrule
Rules only & None & No & Rule predicates & Full corpus \\
Direct & None & No & LLM label & Full corpus and stability subset \\
Claims & None & No & Claim policy & Full corpus and stability subset \\
\midrule
Claims & None & Yes & Claim policy & Stability subset \\
\midrule
Claims & Decompiler & No & Claim policy & Selected cases \\
Claims & Decompiler & Yes & Claim policy & Selected cases \\
Claims & Decompiler and dynamic & No & Claim policy & Selected cases \\
Claims & Decompiler and dynamic & Yes & Claim policy & Selected cases \\
\bottomrule
\end{tabular}}
\end{table*}

\subsection{Evidence Collection}
\subsubsection{Basic Static Evidence}
MARS identifies the executable format and applies the corresponding parser to extract headers, sections, imports or dependencies, available symbols, printable strings, and signature metadata. Additional signals include candidate URLs, domains, IP addresses, registry paths, commands, YARA matches, packer cues, unusual section properties, and trusted code-signing information.

Conceptually related PE and ELF artifacts are grouped under common fields while platform-specific details are retained. For example, Windows imports and ELF dependencies both represent external functionality, whereas registry paths and ELF service or shell commands retain their original forms. Ground-truth labels are excluded from both the evidence bundle and prompts. Local storage paths and experiment identifiers are excluded from prompts. Filenames are replaced with neutral identifiers unless they form part of the intended evidence.

\subsubsection{Decompiler-Assisted Evidence}
The optional decompiler stage runs Ghidra in headless mode to extract function bodies, call relationships, and string references. Function selection excludes thunks, external references, and bodies smaller than 16 bytes. Developer-assigned symbols receive priority over automatically generated names, and code size is ranked in descending order. Selected functions are retained with their available names, addresses, calls, and referenced strings. Limits on the number of functions and the amount of decompiled text are fixed before evaluation.

\subsubsection{Optional Dynamic Evidence}
Dynamic collection executes each sample from a restored virtual machine state for a 60-second observation interval. PE monitoring uses Procmon\footnote{https://learn.microsoft.com/en-us/sysinternals/downloads/procmon} and FakeNet-NG\footnote{https://github.com/mandiant/flare-fakenet-ng} to collect process, file, registry, and emulated network events. ELF monitoring uses strace, auditd, and INetSim\footnote{https://www.inetsim.org} to collect system calls, audit events, and process, file, and network observations.

Collection records distinguish successful observation, execution failure, timeout, and unavailable instrumentation. Missing runtime evidence is marked as unavailable and does not establish behavioral absence. The dynamic study adds runtime observations to static and decompiler evidence for selected cases. Its results describe those cases and do not estimate the benefit of dynamic evidence across the full corpus.

\subsubsection{Unified Evidence Bundle}
Table~\ref{tab:evidence_bundle} lists the main evidence groups. Each modality retains its origin, availability, and platform-specific fields. A bundle is constructed once per sample and evidence setting, then serialized, hashed, and cached. Matched models and decision paths read the same cached bundle, and recorded hashes are checked to verify input identity.

\begin{table*}[t]
\centering
\caption{Main evidence groups in the unified bundle.}
\label{tab:evidence_bundle}
\small
\adjustbox{max width=\textwidth}{%
\begin{tabular}{lll}
\toprule
Evidence group & Representative PE artifacts & Representative ELF artifacts \\
\midrule
File structure & headers, sections, signatures & ELF headers, sections, program headers \\
External functionality & imported APIs and libraries & dynamic dependencies and symbols \\
Strings and indicators & URLs, domains, registry paths, commands & URLs, domains, paths, shell commands \\
Static heuristics & YARA matches, packer cues, section anomalies & YARA matches, packer cues, section anomalies \\
Decompiler evidence & function bodies, calls, referenced strings & function bodies, calls, referenced strings \\
Runtime evidence & processes, files, registry, network events & processes, files, system calls, network events \\
Collection status & parser and execution status & parser and execution status \\
\bottomrule
\end{tabular}}
\end{table*}

Preprocessing applies fixed ordering, deduplication, and field-specific limits on retained item counts before downstream branching. These rules do not depend on ground-truth labels or model outputs. Prompts delimit binary-derived strings, commands, and decompiled text as untrusted data and instruct the model not to follow instructions embedded in these fields.

\subsection{Decision Paths}
\subsubsection{Direct LLM Classification}
The direct path returns one of four labels, benign, suspicious, malicious, or unknown. Its JSON schema also includes confidence and a short justification. The prompt distinguishes observed evidence from inferred capability, cautions against treating isolated APIs, strings, or cryptographic primitives as proof of malicious execution, and permits unknown when evidence is insufficient.

The main runs retain the label and confidence. Justifications for the auxiliary family-information analysis were collected separately for Gemini 2.5 Flash, GLM-5.2, and Gemma4-31B by querying the direct path again with the same prompt and evidence bundle. These texts are therefore separate responses to the same input, not the original justifications accompanying the recorded verdicts.

\subsubsection{Structured Claim Extraction}
The claim path returns a list of atomic propositions. Each claim has an identifier, statement, category, integer severity from 1 to 5, and confidence from 0 to 1. The prompt requires one principal proposition per claim and distinguishes observed artifacts from inferred capabilities. Networking APIs, for example, can support a capability claim but do not establish observed communication. Missing modalities cannot support claims of behavioral absence.

The eleven categories in Table~\ref{tab:claim_taxonomy} provide the interface to the scoring policy. Their weights are prespecified engineering parameters and are not learned from the evaluation corpus. The policy consumes claim categories and severity values. The single-pass configuration scores the extracted list directly.

\begin{table*}[t]
\centering
\caption{Claim taxonomy and prespecified category weights.}
\label{tab:claim_taxonomy}
\small
\adjustbox{max width=\textwidth}{%
\begin{tabular}{lll}
\toprule
Claim category & Representative artifact source & Weight \\
\midrule
Obfuscation or packing & packer cues, unusual sections, loader artifacts & 1.2 \\
Suspicious API usage & sensitive imports, decompiled API calls, injection symbols & 2.2 \\
Network indicator & URLs, domains, IP addresses, network-related strings & 1.0 \\
Persistence candidate & autorun paths, service creation, persistence commands & 2.0 \\
Anti-analysis & debugger checks, timing checks, environment probes & 1.6 \\
Crypto or encoding & cryptographic APIs, encoded strings, encoded data & 0.8 \\
Configuration or embedded data & embedded configuration, mutexes, endpoints & 1.0 \\
Script or command & command lines, shell utilities, script fragments & 1.8 \\
Capability hint & generic functionality not represented above & 1.0 \\
Benign indicator & trusted signatures, known publishers, standard tooling cues & -1.5 \\
Other & residual observations with limited scoring value & 0.5 \\
\bottomrule
\end{tabular}}
\end{table*}

\subsubsection{Second-Pass Claim Support Assessment}
The single-pass and support-assessed configurations use the same extraction and scoring pipeline, with the second pass disabled in the single-pass configuration. The optional second pass receives the evidence bundle and candidate claims and assigns each claim a support state. Supported indicates direct support in the supplied artifacts, hypothesis indicates plausible but incomplete support, and unsupported indicates insufficient support. The implementation stores the supported state as \texttt{verified}. These states are model judgments about the supplied evidence and do not provide independent validation.

The second pass attaches support annotations by claim identifier without adding or removing claims or changing their categories or severities. The support annotations are retained in the output records but are not used by the evaluated scoring policy. For a fixed first-pass claim set, enabling the second pass therefore leaves the score and verdict unchanged.

\begin{figure*}[pos=t]
\centering
\includegraphics[width=\textwidth]{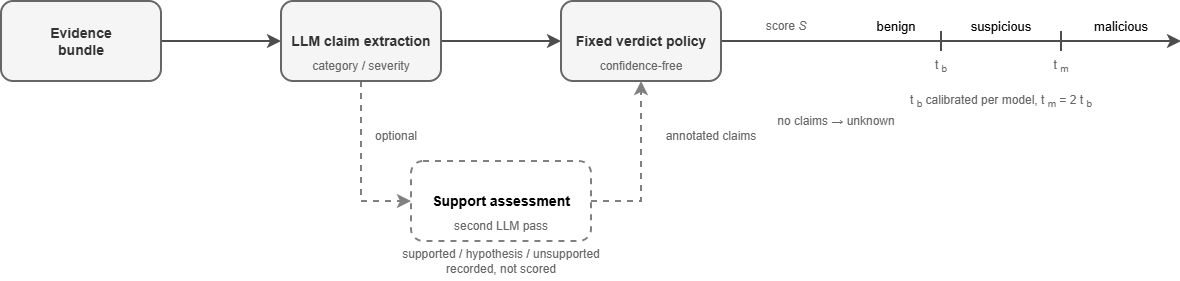}
\caption{Claim processing in MARS. Single-pass extraction supplies claims directly to the scoring policy. The optional second pass attaches support annotations without changing claim categories or severities. The evaluated policy uses category and severity, excludes confidence and support state, and assigns labels using thresholds calibrated or specified before evaluation.}
\label{fig:claim_processing}
\end{figure*}

In the experiments, the two configurations were executed independently, each with a new first-pass extraction. Their claim sets can therefore differ despite identical evidence bundles. Under the evaluated policy, differences in their scores and verdicts arise from independent claim extraction rather than from the support annotations. Figure~\ref{fig:claim_processing} illustrates the optional assessment stage and the scoring policy.

\subsection{Fixed Verdict Policy}
\label{sec:verdict_policy}

For each execution, let $\mathcal{C}(x)$ denote the first-pass claim set generated for sample $x$. Optional support annotations do not alter the categories or severities used by the evaluated scoring policy. Each claim contributes

\begin{equation}
V(c)=w_{\mathrm{type}(c)} \cdot m\bigl(\mathrm{severity}(c)\bigr),
\end{equation}

where $w_{\mathrm{type}(c)}$ is the category weight in Table~\ref{tab:claim_taxonomy}. The severity multiplier $m(s)$ is $0.55$, $0.75$, $1.0$, $1.25$, and $1.5$ for $s=1,\ldots,5$, respectively. The aggregate score is

\begin{equation}
S(x)=\sum_{c \in \mathcal{C}(x)}V(c).
\end{equation}

Confidence is retained in the claim records but excluded from this score. The model influences the score through the extracted claim set, categories, and severity values. A fixed scoring function therefore does not make claim generation deterministic or remove model judgment.

A sample with an empty claim set receives unknown. Otherwise, the label is benign when $S(x)<t_b$, suspicious when $t_b \leq S(x)<t_m$, and malicious when $S(x)\geq t_m$. The benign boundary $t_b$ is selected separately for each model to maximize balanced accuracy on the disjoint 197-sample development set and is then held fixed during evaluation. Table~\ref{tab:model_panel} reports the selected values. The malicious boundary is set to $t_m=2\,t_b$. Because the ground truth distinguishes benign from malicious samples but provides no separate suspicious class, $t_m$ is a specified severity boundary. The malicious label rate describes label use rather than classification correctness.

The rules-only baseline uses the deployed system's deterministic rule predicates without claims or calibrated thresholds. The direct path uses the model's label. Thus, the comparison concerns decision architectures with different verdict mechanisms. Only the claim path uses a threshold calibrated on development data.

The main claim-path comparisons use the policy defined above. The deployed default additionally weights each claim by $\max(0.5,p(c))$, where $p(c)$ is its confidence, and adds deterministic rule contributions before applying fixed thresholds. It also gives supported claims full weight, reduces the contribution of hypotheses, and excludes unsupported claims. The fuller deployed support-assessment procedure includes category-specific evidence requirements, conflict handling, and evidence-reference validation. These components are not included in the evaluated support-assessment configuration, whose results therefore do not characterize that full procedure.

\subsection{LLM Settings and Output Artifacts}
Direct classification and claim extraction use temperature 0.1. Support assessment uses temperature 0.0. Output limits are 4096 tokens for direct classification and 16384 tokens for claim stages. Requests use fixed JSON schemas, up to three automated retries, and a 600-second timeout. Responses that remain invalid are recorded as errors without manual correction. Errors remain in full-corpus verdict denominators. The paired accuracy, family-information, and run-level analyses use complete-case intersections and report their exclusions separately.

\begin{table}[t]
\centering
\caption{Model panel, access interfaces and periods, and calibrated benign boundaries. The malicious boundary is $t_m=2\,t_b$.}
\label{tab:model_panel}
\small
\adjustbox{max width=\linewidth}{%
\begin{tabular}{lllc}
\toprule
Model & Interface & Access period & $t_b$ \\
\midrule
Gemini 2.5 Flash & Google API & 13--14 Jul 2026 & 6.100 \\
DeepSeek-V4-Pro & Ollama Cloud & 13--14 Jul 2026 & 2.575 \\
GLM-5.2 & Ollama Cloud & 13--14 Jul 2026 & 3.025 \\
Gemma4-31B & Ollama Cloud & 26 Aug 2026 & 2.100 \\
GPT-OSS-20B & Ollama Cloud & 26 Aug 2026 & 2.675 \\
Qwen3.5-397B & Ollama Cloud & 27 Aug 2026 & 1.750 \\
\bottomrule
\end{tabular}}
\end{table}

The model identifiers are floating tags rather than dated snapshots. Ollama Cloud does not expose the served quantization or hardware to the client. These constraints limit exact reproduction, and low temperature does not guarantee identical outputs. Gemma3-27B and Qwen3-Coder-480B were excluded from the six-model panel after provider withdrawal. Their archived claims remain available for score and label replay without model access.

Run artifacts include the evidence bundle and hash, direct label and confidence, extracted claims and support annotations, applicable rule contributions and scores, final labels, errors, token counts, stage runtimes, prompt versions, and model settings. These records support reconstruction of the downstream scoring process. The analysis scripts were developed with LLM assistance, and we verified the reported values against the stored artifacts.

\subsection{Scoring Policy Sensitivity}
Sensitivity analyses reuse stored claims to vary category weights and the benign boundary without additional LLM calls. Joint weight sensitivity is assessed with 200 random perturbations of up to $\pm 50\%$ per weight, refitting $t_b$ on the development set after each draw. A separate analysis varies each weight from $0.5\times$ to $1.5\times$ its prespecified value while holding the other weights fixed.

For threshold sensitivity, an oracle analysis selects $t_b$ to maximize balanced accuracy directly on the evaluation corpus. This explicitly uses evaluation labels and serves only as a diagnostic of threshold dependence. It does not determine the threshold used for the main comparison. Two-fold cross-validation within the development set fits the threshold on one fold and evaluates it on the other, with the roles exchanged. This analysis examines threshold generalization within the development set alongside the performance change when transferring the calibrated threshold to the evaluation corpus.

\section{Evaluation}
\subsection{Dataset Construction and Selection}
Malicious PE and ELF samples were obtained from MalwareBazaar\footnote{https://bazaar.abuse.ch} using family tags and a retrieval cap per family. The retained samples span 28 family labels. These annotations are used only in the auxiliary family-identification analysis, whereas verdict evaluation uses binary benign and malicious labels assigned before MARS processing. Benign PE samples include signed vendor applications, unsigned open-source tools, UPX-packed legitimate software, and installers. Benign ELF samples were extracted from Debian packages covering multiple instruction set architectures.

The evaluation corpus contains 1,195 binaries. The PE set has 297 benign and 300 malicious samples, and the ELF set has 300 benign and 298 malicious samples. A separate development set contains 197 binaries, comprising 43 benign and 52 malicious PE samples and 50 benign and 52 malicious ELF samples. It is used for threshold calibration.

Near-duplicates are grouped separately within each platform and class using ppdeep similarity at a threshold of 70 on its 0--100 scale. Samples are processed in a fixed order and assigned to the first cluster whose representative meets the threshold. A sample with no match becomes the representative of a new cluster. This procedure produces 1,001 clusters, comprising 530 PE and 471 ELF clusters. Import hashes are excluded from clustering because they over-group .NET and packed binaries. The development set follows the same construction procedure, and cross-deduplication between development and evaluation samples uses ppdeep at a threshold of 85. Cluster identifiers define resampling units and cross-validation groups.

The corpus is neither family-balanced nor time-balanced, and its class proportions do not represent deployment prevalence. The benign collection includes several sources of false alarms but does not exhaust difficult categories such as potentially unwanted programs and administrative tools.

\subsection{Experimental Design and Metrics}
The main comparison uses basic static evidence with the configurations in Table~\ref{tab:processing_configurations}. Each sample is processed once by the rules-only baseline and once per model by direct classification and single-pass claim processing. The six-model panel includes Gemini 2.5 Flash~\cite{comanici2025gemini} and is specified in Table~\ref{tab:model_panel}. Matched paths use identical cached evidence. The claim policy and development-calibrated thresholds remain fixed throughout the main comparison.

An exploratory follow-up with Gemini 2.5 Flash compares static-plus-decompiler evidence with the same evidence augmented by runtime observations. It uses 20 selected samples, comprising three benign controls and seven malicious samples per platform. Both single-pass and support-assessed claim paths are evaluated. The retained records do not specify a reproducible selection rule, so these results are descriptive and do not provide a representative estimate of the benefit of dynamic evidence.

The repeated-run study samples one binary per near-duplicate cluster from the evaluation corpus using a fixed random seed, with 15 benign and 15 malicious samples per platform. Direct, single-pass, and support-assessed paths are each executed ten times with Gemini 2.5 Flash, DeepSeek-V4-Pro, and GLM-5.2. Support assessment is not evaluated on the full corpus.

An alert is either suspicious or malicious. The benign exact-label rate is the proportion of benign samples labeled benign, and benign escalation is the proportion assigned an alert. Malicious alert recall is the proportion of malicious samples assigned an alert. Unknown outputs and processing errors count as incorrect outcomes for accuracy and as non-alerts, but not as benign escalations. Overall accuracy combines correct benign labels and alerts on malicious samples. Balanced accuracy is the mean of the benign exact-label rate and malicious alert recall. The malicious label rate measures use of that label without treating the suspicious--malicious distinction as ground truth.

Errors remain in full-corpus denominators. Analyses requiring complete records specify their exclusions. Individual rate intervals use the Wilson method~\cite{wilson1927probable}. For paired accuracy differences, we resample near-duplicate clusters with replacement 5,000 times using a random seed of 42. The 95\% confidence interval is defined by the 2.5th and 97.5th percentiles of the bootstrap differences. Sample-level McNemar tests use a two-sided exact binomial test on discordant pairs and are supplementary because they do not adjust for dependence within clusters. Stability is measured by agreement across all ten runs, either on the exact four-class label or on the binary alert outcome. Runtime is summarized by the median.

\subsection{Verdict Results}
\subsubsection{Shallow-Metadata Baselines}
Table~\ref{tab:difficulty_floor} compares MARS with a majority-class baseline, an architecture-only baseline using one-hot encoding, and a metadata classifier using architecture, log file size, byte entropy, packer status, and signature status. Both trained baselines respect cluster boundaries in cross-validation. The metadata classifier uses histogram-based gradient boosting with 300 iterations and a fixed seed, evaluated separately by platform using five folds. These features contain no decoded instructions, string content, import names, or runtime observations.

\begin{table*}[t]
\centering
\caption{Shallow-metadata baselines. Direct and single-pass rows report means over six models. MARS uses no model fine-tuning, and the single-pass path calibrates a per-model threshold on a disjoint development set.}
\label{tab:difficulty_floor}
\small
\adjustbox{max width=\textwidth}{%
\begin{tabular}{llcccc}
\toprule
Platform & Condition & Fitted to corpus & Malicious alert recall & Benign exact-label rate & Balanced accuracy \\
\midrule
PE & Majority class & -- & 100.0\% & 0.0\% & 50.0\% \\
PE & Architecture one-hot & Yes & 85.3\% & 88.6\% & 86.9\% \\
PE & Architecture + four metadata features & Yes & 93.7\% & 93.6\% & 93.6\% \\
PE & MARS rules only & No & 50.7\% & 32.7\% & 41.7\% \\
PE & MARS direct (mean of 6) & No & 90.8\% & 93.9\% & 92.4\% \\
PE & MARS single-pass (mean of 6) & No & 80.2\% & 77.7\% & 79.0\% \\
\midrule
ELF & Majority class & -- & 0.0\% & 100.0\% & 50.0\% \\
ELF & Architecture one-hot & Yes & 80.2\% & 88.3\% & 84.3\% \\
ELF & Architecture + four metadata features & Yes & 84.2\% & 93.3\% & 88.8\% \\
ELF & MARS rules only & No & 0.0\% & 16.3\% & 8.2\% \\
ELF & MARS direct (mean of 6) & No & 82.6\% & 99.3\% & 90.9\% \\
ELF & MARS single-pass (mean of 6) & No & 67.7\% & 97.3\% & 82.5\% \\
\bottomrule
\end{tabular}}
\end{table*}

Architecture alone reaches 86.9\% balanced accuracy on PE and 84.3\% on ELF. Adding the four metadata features raises these values to 93.6\% and 88.8\%. The metadata classifier exceeds mean single-pass performance on both platforms and mean direct performance on PE. It is fitted on training folds from this corpus, whereas MARS uses no model fine-tuning and calibrates claim thresholds on separate development data. The results establish substantial class separation from shallow features and limit what absolute verdict accuracy can demonstrate about behavioral analysis.

\subsubsection{Results by Platform}
Table~\ref{tab:verdict_results} reports full-corpus outcomes. Direct classification has higher malicious alert recall in ten of twelve platform and model combinations. The exceptions are Gemma4-31B on PE and DeepSeek-V4-Pro on ELF.

\begin{table*}[t]
\centering
\caption{Verdict results on the full evaluation corpus with basic static evidence, over 297 benign and 300 malicious PE samples and 300 benign and 298 malicious ELF samples.}
\label{tab:verdict_results}
\adjustbox{max width=\textwidth}{
\begin{tabular}{llccccc}
\toprule
Decision path & Model & Benign exact & Benign escalation & Malicious alert & Malicious label & Errors \\
\midrule
\multicolumn{7}{l}{PE} \\
\midrule
Rules only (baseline) & -- & 97/297 (32.7\%) & 72/297 (24.2\%) & 152/300 (50.7\%) & 24/300 (8.0\%) & 0 \\
\midrule
Direct & Gemini 2.5 Flash & 242/297 (81.5\%) & 55/297 (18.5\%) & 298/300 (99.3\%) & 290/300 (96.7\%) & 1 \\
Single-pass & Gemini 2.5 Flash & 259/297 (87.2\%) & 38/297 (12.8\%) & 173/300 (57.7\%) & 63/300 (21.0\%) & 1 \\
Direct & DeepSeek-V4-Pro & 289/297 (97.3\%) & 7/297 (2.4\%) & 255/300 (85.0\%) & 200/300 (66.7\%) & 2 \\
Single-pass & DeepSeek-V4-Pro & 241/297 (81.1\%) & 56/297 (18.9\%) & 246/300 (82.0\%) & 149/300 (49.7\%) & 2 \\
Direct & GLM-5.2 & 294/297 (99.0\%) & 3/297 (1.0\%) & 275/300 (91.7\%) & 242/300 (80.7\%) & 2 \\
Single-pass & GLM-5.2 & 247/297 (83.2\%) & 50/297 (16.8\%) & 241/300 (80.3\%) & 143/300 (47.7\%) & 2 \\
Direct & Gemma4-31B & 292/297 (98.3\%) & 5/297 (1.7\%) & 230/300 (76.7\%) & 224/300 (74.7\%) & 2 \\
Single-pass & Gemma4-31B & 215/297 (72.4\%) & 82/297 (27.6\%) & 290/300 (96.7\%) & 126/300 (42.0\%) & 3 \\
Direct & GPT-OSS-20B & 264/297 (88.9\%) & 31/297 (10.4\%) & 286/300 (95.3\%) & 280/300 (93.3\%) & 2 \\
Single-pass & GPT-OSS-20B & 190/297 (64.0\%) & 105/297 (35.4\%) & 218/300 (72.7\%) & 92/300 (30.7\%) & 3 \\
Direct & Qwen3.5-397B & 293/297 (98.7\%) & 1/297 (0.3\%) & 290/300 (96.7\%) & 289/300 (96.3\%) & 13 \\
Single-pass & Qwen3.5-397B & 233/297 (78.5\%) & 64/297 (21.5\%) & 276/300 (92.0\%) & 191/300 (63.7\%) & 2 \\
\midrule
\multicolumn{7}{l}{ELF} \\
\midrule
Rules only (baseline) & -- & 49/300 (16.3\%) & 40/300 (13.3\%) & 0/298 (0.0\%) & 0/298 (0.0\%) & 0 \\
\midrule
Direct & Gemini 2.5 Flash & 290/300 (96.7\%) & 10/300 (3.3\%) & 283/298 (95.0\%) & 265/298 (88.9\%) & 10 \\
Single-pass & Gemini 2.5 Flash & 295/300 (98.3\%) & 5/300 (1.7\%) & 133/298 (44.6\%) & 49/298 (16.4\%) & 17 \\
Direct & DeepSeek-V4-Pro & 300/300 (100.0\%) & 0/300 (0.0\%) & 218/298 (73.2\%) & 188/298 (63.1\%) & 10 \\
Single-pass & DeepSeek-V4-Pro & 297/300 (99.0\%) & 3/300 (1.0\%) & 234/298 (78.5\%) & 145/298 (48.7\%) & 10 \\
Direct & GLM-5.2 & 300/300 (100.0\%) & 0/300 (0.0\%) & 276/298 (92.6\%) & 252/298 (84.6\%) & 10 \\
Single-pass & GLM-5.2 & 299/300 (99.7\%) & 1/300 (0.3\%) & 222/298 (74.5\%) & 153/298 (51.3\%) & 10 \\
Direct & Gemma4-31B & 298/300 (99.3\%) & 2/300 (0.7\%) & 256/298 (85.9\%) & 235/298 (78.9\%) & 10 \\
Single-pass & Gemma4-31B & 297/300 (99.0\%) & 3/300 (1.0\%) & 205/298 (68.8\%) & 99/298 (33.2\%) & 10 \\
Direct & GPT-OSS-20B & 299/300 (99.7\%) & 1/300 (0.3\%) & 177/298 (59.4\%) & 158/298 (53.0\%) & 10 \\
Single-pass & GPT-OSS-20B & 269/300 (89.7\%) & 31/300 (10.3\%) & 168/298 (56.4\%) & 57/298 (19.1\%) & 10 \\
Direct & Qwen3.5-397B & 300/300 (100.0\%) & 0/300 (0.0\%) & 267/298 (89.6\%) & 266/298 (89.3\%) & 28 \\
Single-pass & Qwen3.5-397B & 295/300 (98.3\%) & 5/300 (1.7\%) & 248/298 (83.2\%) & 162/298 (54.4\%) & 10 \\
\bottomrule
\end{tabular}}
\end{table*}

On PE, direct classification reaches benign exact-label rates of 81.5--99.0\%, compared with 64.0--87.2\% for single-pass processing. Gemma4-31B gains recall through claims, from 76.7\% to 96.7\%, while its benign exact-label rate falls from 98.3\% to 72.4\%. Gemini 2.5 Flash shows the reverse trade-off. Claims improve its benign exact-label rate from 81.5\% to 87.2\% but reduce recall from 99.3\% to 57.7\%.

\begin{figure*}[t]
\centering
\includegraphics[width=\textwidth]{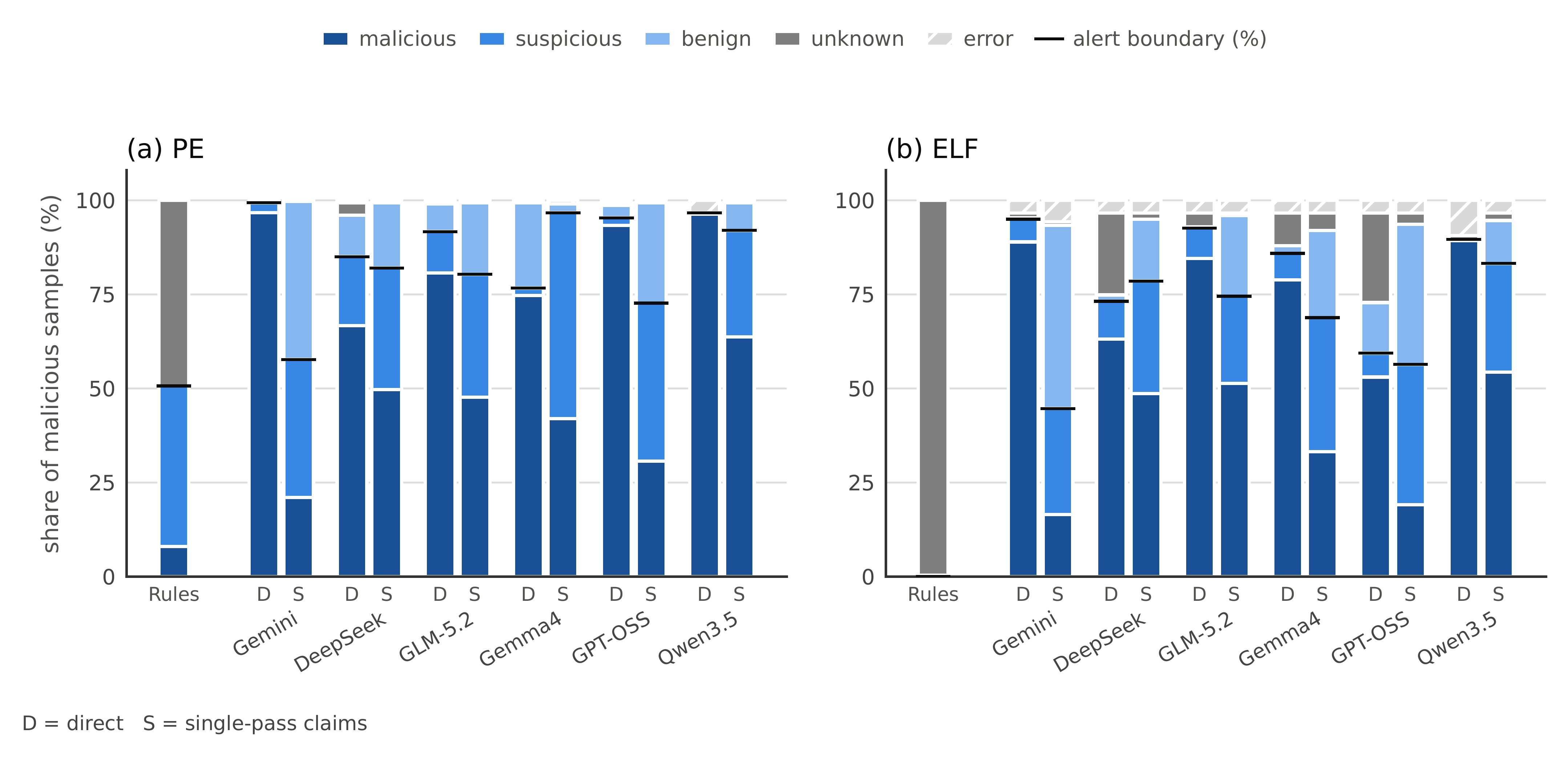}
\caption{Label distribution on malicious samples. Suspicious and malicious count as alerts. Unknown outputs and errors count as non-alerts and remain in the denominators. The rules-only baseline is shown once per platform.}
\label{fig:verdict_distribution}
\end{figure*}

On ELF, benign exact-label rates are 96.7--100.0\% for direct classification and 89.7--99.7\% for single-pass processing. Direct recall ranges from 59.4\% to 95.0\%, compared with 44.6--83.2\% for claims. DeepSeek-V4-Pro is the exception, with recall of 78.5\% under claims and 73.2\% under direct classification. Figure~\ref{fig:verdict_distribution} shows label use across both platforms. Direct classification generally assigns malicious more often, whereas the claim policy assigns suspicious to a larger share of alerted samples.

The rules-only baseline reaches malicious recall of 50.7\% on PE and 0.0\% on ELF, with benign exact-label rates of 32.7\% and 16.3\%. These comparisons do not isolate the contribution of claims because the rules-only and claim paths use different verdict policies.

\subsubsection{ELF Rule Coverage}
The ELF rule layer assigns suspicious at a score of 3.0 and malicious at 7.0. Each YARA match contributes 1.5. The remaining escalation term is a 1.5 bonus requiring at least two imported symbols in the process-injection, anti-analysis, or keylogging categories. The unsigned-binary penalty is disabled for ELF. Recomputing the rules over all 598 cached bundles reproduces every recorded label. No malicious sample exceeds 1.5, whereas 40 benign samples reach 3.0. Thus, the alert boundary is reachable, but none of the malicious samples in this corpus reaches it.

Of the 298 malicious ELF samples, 251 have no YARA match and 47 have one. None of the latter also receives the symbol bonus. The ELF taxonomy includes a process-manipulation category that the bonus predicate does not recognize. The bonus instead fires on 70 benign and 13 malicious samples through the remaining eligible categories. YARA is active, with matches in 97 bundles, but only three rules match and the repository contains no ELF-specific rule. These findings identify limited coverage of this corpus by the implemented rules. The two LLM paths do not use this bonus, and the result does not establish an inherent lack of information in ELF binaries.

\subsubsection{Paired Comparison}
Table~\ref{tab:paired_stats} compares direct and single-pass accuracy on samples with usable outputs from both paths. The matched sample count ranges from 1,154 to 1,183. Qwen3.5-397B has the smallest intersection because 29 of its 41 direct errors occur where claim extraction succeeds. An error in either path excludes that sample from both sides of this paired comparison.

\begin{table*}[t]
\centering
\caption{Paired accuracy on complete-case intersections. Differences use cluster-bootstrap 95\% CIs. Exact McNemar $p$-values are sample-level results without adjustment for within-cluster dependence. Accuracy values are percentages and differences are percentage points.}
\label{tab:paired_stats}
\small
\adjustbox{max width=\textwidth}{
\begin{tabular}{lcccccc}
\toprule
Model & $n$ & Direct accuracy [95\% CI] & Single-pass accuracy [95\% CI] & $\Delta$ (pp) & Bootstrap 95\% CI & McNemar $p$ \\
\midrule
Gemini 2.5 Flash & 1177 & 94.0 [92.5, 95.2] & 73.1 [70.5, 75.5] & $20.9$ & $[16.6, 26.0]$ & $3.1\times10^{-45}$ \\
DeepSeek-V4-Pro & 1183 & 89.8 [87.9, 91.4] & 86.1 [84.0, 87.9] & $3.7$ & $[1.6, 6.0]$ & $5.8\times10^{-4}$ \\
GLM-5.2 & 1183 & 96.8 [95.6, 97.7] & 85.3 [83.2, 87.2] & $11.5$ & $[8.8, 14.6]$ & $7.7\times10^{-35}$ \\
Gemma4-31B & 1182 & 90.9 [89.2, 92.5] & 85.2 [83.1, 87.1] & $5.8$ & $[3.1, 8.5]$ & $3.5\times10^{-6}$ \\
GPT-OSS-20B & 1182 & 86.7 [84.7, 88.5] & 71.5 [68.8, 74.0] & $15.2$ & $[12.7, 17.8]$ & $2.0\times10^{-33}$ \\
Qwen3.5-397B & 1154 & 99.7 [99.1, 99.9] & 90.6 [88.8, 92.2] & $9.0$ & $[7.3, 11.0]$ & $3.6\times10^{-29}$ \\
\bottomrule
\end{tabular}}
\end{table*}

Direct classification leads by 3.7--20.9 percentage points, with all six cluster-bootstrap intervals above zero. It also leads when errors remain in the denominator, as reported in Table~\ref{tab:hard_subset}. The ordering persists in the diagnostic analysis that selects the claim threshold using evaluation labels. It is specific to the six-model panel. Archived results for the withdrawn Gemma3-27B instead favor claims by 16.3 points, associated with substantial benign escalation under direct classification.

\subsection{Checks on the Verdict Comparison}
\subsubsection{Fixed Architecture Strata}
Architecture is strongly associated with class in this corpus. Malicious samples account for 88.3\% of PE x86 binaries but 14.3\% of PE x64 binaries, and for 80.2\% of ELF x86-32 binaries but 17.4\% of ELF x86-64 binaries. We therefore compare the paths within these four architecture strata. Other strata have insufficient benign coverage for a reliable comparison.

Direct classification retains higher mean balanced accuracy in all four strata, with margins of 12.1 points on PE x64, 24.3 on PE x86, 7.6 on ELF x86-64, and 4.4 on ELF x86-32. It leads in 21 of 24 model and stratum combinations. Holding architecture constant preserves the overall ordering, although other metadata cues remain available within each stratum.

\subsubsection{Metadata-Hard Subset}
An exploratory analysis selects samples on which the metadata classifier is least confident. For each sample, we compute the absolute distance of its out-of-fold predicted probability from 0.5. The lowest quartile within each platform, including ties at the cutoff, yields 150 samples per platform and 300 in total. The selected subset is also balanced by class. Metadata accuracy is 71.0\% on this subset, compared with 91.2\% overall and 98.0\% on its complement.

\begin{table*}[t]
\centering
\caption{Accuracy on the full corpus and the metadata-hard subset, with errors retained in the denominators. Accuracy values are percentages and differences are percentage points.}
\label{tab:hard_subset}
\adjustbox{max width=\textwidth}{%
\begin{tabular}{lccccc}
\toprule
& \multicolumn{2}{c}{Full corpus (1,195)} & \multicolumn{2}{c}{Metadata-hard (300)} & \\
\cmidrule(lr){2-3}\cmidrule(lr){4-5}
Model & Direct & Single-pass & Direct & Single-pass & $\Delta$ on hard \\
\midrule
Gemini 2.5 Flash & 93.1 & 72.0 & 87.3 & 66.7 & $+20.7$ \\
DeepSeek-V4-Pro & 88.9 & 85.2 & 88.7 & 82.0 & $+6.7$ \\
GLM-5.2 & 95.8 & 84.4 & 94.0 & 82.3 & $+11.7$ \\
Gemma4-31B & 90.0 & 84.3 & 89.3 & 81.0 & $+8.3$ \\
GPT-OSS-20B & 85.9 & 70.7 & 84.0 & 67.3 & $+16.7$ \\
Qwen3.5-397B & 96.2 & 88.0 & 94.0 & 83.3 & $+10.7$ \\
\bottomrule
\end{tabular}}
\end{table*}

Direct classification remains more accurate on all six models. The median accuracy gap is 11.2 percentage points on the subset and 9.8 on the full corpus. The result supports the ordering on samples with weaker separation under this particular metadata classifier. It does not establish difficulty independent of that classifier or demonstrate behavioral reasoning.

\subsubsection{Collector Failures}
Eleven malicious samples, ten ELF and one PE, fail in both LLM paths for every model. The ten ELF evidence bundles are at least 4.9 MB in size, with a maximum of 25.2 MB. The PE evidence bundle is 2.9 MB. The collector limits item counts but permits individual strings to span millions of characters and imposes no overall serialized-size limit. Excluding these samples, the remaining request error rate is 0.34\%.

These failures remain in the full-corpus metrics. A sensitivity analysis excludes them from both paths without new model calls, reducing the malicious denominators from 298 to 288 on ELF and from 300 to 299 on PE. Benign metrics are unchanged. Direct classification still has higher recall in ten of twelve platform and model combinations and higher overall accuracy for all six models. The largest change in an accuracy gap is 0.20 percentage points. Thus, the collector defect affects absolute performance but does not explain the observed ordering.

\subsection{Family Information in Retained Outputs}
We measure family identification from retained representations as an auxiliary probe, not as a MARS output. Targets are MalwareBazaar family annotations, evaluated separately by platform. The main comparison requires all evaluated representations to be available for each sample, including claim sets from all six models, direct justifications from three models, and both verdict encodings. After matching, we exclude families with fewer than five samples within each platform. This leaves 285 PE samples from 16 families and 260 ELF samples from 10 families.

The representations are one-hot verdicts, static metadata, evidence-bundle text, claim statements and types, and direct justifications. Metadata includes size, entropy, signature status, packer status, architecture, and import-hash sharing. Family names are removed from the text representations. Evidence text, claims, and justifications use TF-IDF features with identical settings, comprising unigrams and bigrams, a maximum of 4,000 features, a minimum document frequency of two, and sublinear term-frequency scaling. Static metadata is standardized using StandardScaler. Multinomial logistic regression uses $C=5.0$ and a maximum of 2,000 iterations. Evaluation uses five-fold GroupKFold cross-validation with near-duplicate clusters as groups, and vectorizers and scalers are fitted within training folds. Justifications are available for Gemini 2.5 Flash, GLM-5.2, and Gemma4-31B from separate direct calls using the original prompt and bundle. They are not the original responses accompanying the recorded verdicts.

\begin{table*}[t]
\centering
\caption{Family identification on the main matched subsets of 285 PE and 260 ELF samples.}
\label{tab:class_information}
\small
\adjustbox{max width=\textwidth}{%
\begin{tabular}{lllcc}
\toprule
Platform & Representation & Uses LLM & Top-1 accuracy & Macro-F1 \\
\midrule
PE & Majority family & -- & 10.5\% & -- \\
PE & Direct verdict only & Yes & 3.9--10.2\% & $\leq 0.052$ \\
PE & Claim-path verdict only & Yes & 8.4--13.7\% & $\leq 0.050$ \\
PE & Static metadata & No & 41.8\% & 0.336 \\
PE & Claims (Gemini 2.5 Flash) & Yes & 54.4\% & 0.499 \\
PE & Claims (DeepSeek-V4-Pro) & Yes & 53.7\% & 0.498 \\
PE & Claims (GLM-5.2) & Yes & 57.9\% & 0.531 \\
PE & Claims (Gemma4-31B) & Yes & 56.5\% & 0.521 \\
PE & Claims (GPT-OSS-20B) & Yes & 44.2\% & 0.407 \\
PE & Claims (Qwen3.5-397B) & Yes & 51.2\% & 0.472 \\
PE & Direct justification (Gemini 2.5 Flash) & Yes & 53.3\% & 0.485 \\
PE & Direct justification (GLM-5.2) & Yes & 54.0\% & 0.503 \\
PE & Direct justification (Gemma4-31B) & Yes & 55.4\% & 0.509 \\
PE & Evidence-bundle text & No & 61.1\% & 0.591 \\
\midrule
ELF & Majority family & -- & 15.4\% & -- \\
ELF & Direct verdict only & Yes & 5.4--14.6\% & $\leq 0.076$ \\
ELF & Claim-path verdict only & Yes & 8.5--18.5\% & $\leq 0.085$ \\
ELF & Static metadata & No & 44.6\% & 0.426 \\
ELF & Claims (Gemini 2.5 Flash) & Yes & 76.5\% & 0.661 \\
ELF & Claims (DeepSeek-V4-Pro) & Yes & 76.9\% & 0.681 \\
ELF & Claims (GLM-5.2) & Yes & 81.9\% & 0.707 \\
ELF & Claims (Gemma4-31B) & Yes & 81.2\% & 0.732 \\
ELF & Claims (GPT-OSS-20B) & Yes & 72.3\% & 0.607 \\
ELF & Claims (Qwen3.5-397B) & Yes & 78.1\% & 0.682 \\
ELF & Direct justification (Gemini 2.5 Flash) & Yes & 71.9\% & 0.638 \\
ELF & Direct justification (GLM-5.2) & Yes & 80.4\% & 0.725 \\
ELF & Direct justification (Gemma4-31B) & Yes & 81.9\% & 0.727 \\
ELF & Evidence-bundle text & No & 83.1\% & 0.740 \\
\bottomrule
\end{tabular}}
\end{table*}

Table~\ref{tab:class_information} summarizes the main comparison. Verdict-only accuracy remains near or below the majority-family baseline, with macro-F1 below 0.09. Claim accuracy exceeds the metadata baseline on both platforms, ranging from 44.2\% to 57.9\% on PE and 72.3\% to 81.9\% on ELF. However, evidence-bundle text has higher point accuracy than every claim representation, reaching 61.1\% on PE and 83.1\% on ELF. Differences between claims and the corresponding direct justifications range from $-0.7$ to $+4.6$ percentage points.

A separate field ablation uses 300 PE and 296 ELF samples. Removing dedicated URL, domain, and address fields changes evidence-text accuracy from 59.7\% to 58.0\% on PE and from 82.1\% to 82.8\% on ELF. Strings alone reach 56.0\% and 80.7\%, and import names alone reach 30.0\% and 48.0\%. The small effect of removing dedicated network fields does not exclude reliance on identifying material retained elsewhere in the strings.

An additional comparison uses capa, which processes 290 of 300 PE samples. This comparison requires successfully parsed capa output and available claim sets, without requiring direct justifications. We apply the minimum family size of five after constructing this intersection, leaving 287 PE samples from 17 families. On this subset, accuracy is 63.1\% for capa and 61.0\% for evidence-bundle text. Gemma4-31B yields the highest accuracy among the claim representations at 54.4\%, below both alternatives on the same subset. On ELF, capa processes 68 of 298 samples. Matching leaves 62 samples, and the family-size filter leaves 55, so no ELF classification comparison is reported. Unsupported architecture accounts for 168 processing failures in this experiment.

The probe shows that claims retain family-discriminative content beyond the verdict, but it does not demonstrate greater retention than the evidence text or a consistent advantage over free-text justifications. These are results for a particular classifier and task, not an information-theoretic measure of loss. Family annotations were not independently verified, and their distribution reflects the collection cap. The broader coverage relative to capa also belongs to the shared evidence collector and is available to the text baseline without an LLM.

\subsection{Processing Cost and Additional Evidence}
\subsubsection{Full-Corpus Processing Cost}
Table~\ref{tab:main_panel_cost} reports median latency and token counts over 1,195 samples, together with estimated token costs. Dollar figures are estimates rather than billed amounts. For each sample, the estimate is computed as $(n_{\mathrm{in}}r_{\mathrm{in}} + n_{\mathrm{out}}r_{\mathrm{out}})/10^6$, where $n_{\mathrm{in}}$ and $n_{\mathrm{out}}$ are recorded token counts and the rates are expressed in USD per million tokens. The fixed assumed input and output rates are 0.30 and 2.50 for Gemini 2.5 Flash and 0.40 and 1.20 for every Ollama Cloud model. Per-sample estimates are averaged and scaled to 1,000 samples. Because the five Ollama Cloud models share the same assumed rates, differences in their estimated costs reflect token usage rather than model-specific vendor prices.

\begin{table*}[t]
\centering
\caption{Median latency and token counts, with mean estimated token costs per 1,000 samples under the fixed assumed rates stated in the text. Cost ratios compare single-pass processing with direct classification.}
\label{tab:main_panel_cost}
\adjustbox{max width=\textwidth}{%
\begin{tabular}{lcccccc}
\toprule
Model & Path & Latency (s) & Input tokens & Output tokens & Estimated cost per 1,000 (USD) & Cost ratio \\
\midrule
Gemini 2.5 Flash & Direct & 11.8 & 6474 & 289 & 3.39 & \\
Gemini 2.5 Flash & Single-pass & 15.8 & 7005 & 746 & 4.78 & $1.41\times$ \\
DeepSeek-V4-Pro & Direct & 7.9 & 5513 & 705 & 3.70 & \\
DeepSeek-V4-Pro & Single-pass & 16.2 & 6018 & 1731 & 5.22 & $1.41\times$ \\
GLM-5.2 & Direct & 5.9 & 5358 & 593 & 3.43 & \\
GLM-5.2 & Single-pass & 10.3 & 5842 & 1246 & 4.41 & $1.28\times$ \\
Gemma4-31B & Direct & 2.1 & 6503 & 248 & 3.50 & \\
Gemma4-31B & Single-pass & 3.0 & 7331 & 530 & 4.18 & $1.20\times$ \\
GPT-OSS-20B & Direct & 3.2 & 5337 & 406 & 3.22 & \\
GPT-OSS-20B & Single-pass & 7.3 & 6116 & 1114 & 4.50 & $1.40\times$ \\
Qwen3.5-397B & Direct & 29.2 & 6195 & 2519 & 6.06 & \\
Qwen3.5-397B & Single-pass & 18.3 & 7015 & 1580 & 5.80 & $0.96\times$ \\
\bottomrule
\end{tabular}}
\end{table*}

Under the assumed rates, claim processing has an estimated token cost 1.20--1.41 times that of direct classification and a median latency 1.3--2.3 times as high on five models. On these models, median input token counts are 8--15\% higher, and median output token counts are 2.1--2.7 times those of direct classification. Qwen3.5-397B reverses the pattern, with 2,519 median output tokens for direct classification and 1,580 for claims. Its claim path has lower median latency and a slightly lower estimated token cost. These estimates characterize token usage under the assumed rates and do not establish differences in actual deployment expenditure.

\subsubsection{Dynamic Evidence on Selected Cases}
Dynamic evidence is evaluated with Gemini 2.5 Flash on 20 selected cases, with three benign and seven malicious samples per platform. Static-plus-decompiler evidence provides the control for both claim paths. On PE, malicious alerts increase from 5/7 to 7/7 for single-pass processing and from 4/7 to 6/7 in the support-assessed configuration. All three benign samples remain benign. Malicious labels increase from 0/7 to 5/7 and from 1/7 to 3/7, respectively.

On ELF, support-assessed alerts increase from 4/7 to 6/7, while single-pass alerts decrease from 6/7 to 5/7. The single-pass malicious label count remains 2/7, although the labeled samples change. No processing errors are reported for the dynamic configurations, whose median runtimes range from 174.1 to 233.3 seconds. These selected cases show mixed effects and do not support a corpus-wide estimate of benefit.

\subsection{Repeated-Run Stability}
Table~\ref{tab:stability_results_revised} reports agreement across ten runs using the same verdict policies as the main comparison. One malicious ELF sample fails in every run because of the collector defect, leaving 30 PE and 29 ELF samples.

\begin{table*}[t]
\centering
\caption{Agreement across ten runs on the matched three-model subset. The claim configurations use independent first-pass extractions, and support annotations do not affect the evaluated score. One malicious ELF sample is excluded because the collector defect prevents a usable output in every run.}
\label{tab:stability_results_revised}
\adjustbox{max width=\textwidth}{
\begin{tabular}{llcccc}
\toprule
Decision path & Model & Exact verdict consistency & Benign exact consistency & Malicious exact consistency & Coarse alert consistency \\
\midrule
\multicolumn{6}{l}{PE (15 benign, 15 malicious)} \\
Direct & Gemini 2.5 Flash & 19/30 (63.3\%) & 6/15 (40.0\%) & 13/15 (86.7\%) & 21/30 (70.0\%) \\
Direct & DeepSeek-V4-Pro & 19/30 (63.3\%) & 14/15 (93.3\%) & 5/15 (33.3\%) & 26/30 (86.7\%) \\
Direct & GLM-5.2 & 21/30 (70.0\%) & 15/15 (100.0\%) & 6/15 (40.0\%) & 27/30 (90.0\%) \\
Single-pass & Gemini 2.5 Flash & 15/30 (50.0\%) & 9/15 (60.0\%) & 6/15 (40.0\%) & 20/30 (66.7\%) \\
Single-pass & DeepSeek-V4-Pro & 20/30 (66.7\%) & 12/15 (80.0\%) & 8/15 (53.3\%) & 20/30 (66.7\%) \\
Single-pass & GLM-5.2 & 16/30 (53.3\%) & 13/15 (86.7\%) & 3/15 (20.0\%) & 19/30 (63.3\%) \\
Support-assessed & Gemini 2.5 Flash & 11/30 (36.7\%) & 8/15 (53.3\%) & 3/15 (20.0\%) & 16/30 (53.3\%) \\
Support-assessed & DeepSeek-V4-Pro & 16/30 (53.3\%) & 10/15 (66.7\%) & 6/15 (40.0\%) & 17/30 (56.7\%) \\
Support-assessed & GLM-5.2 & 21/30 (70.0\%) & 13/15 (86.7\%) & 8/15 (53.3\%) & 23/30 (76.7\%) \\
\midrule
\multicolumn{6}{l}{ELF (15 benign, 14 malicious)} \\
Direct & Gemini 2.5 Flash & 27/29 (93.1\%) & 15/15 (100.0\%) & 12/14 (85.7\%) & 29/29 (100.0\%) \\
Direct & DeepSeek-V4-Pro & 22/29 (75.9\%) & 15/15 (100.0\%) & 7/14 (50.0\%) & 25/29 (86.2\%) \\
Direct & GLM-5.2 & 24/29 (82.8\%) & 15/15 (100.0\%) & 9/14 (64.3\%) & 29/29 (100.0\%) \\
Single-pass & Gemini 2.5 Flash & 21/29 (72.4\%) & 15/15 (100.0\%) & 6/14 (42.9\%) & 26/29 (89.7\%) \\
Single-pass & DeepSeek-V4-Pro & 19/29 (65.5\%) & 14/15 (93.3\%) & 5/14 (35.7\%) & 24/29 (82.8\%) \\
Single-pass & GLM-5.2 & 25/29 (86.2\%) & 15/15 (100.0\%) & 10/14 (71.4\%) & 28/29 (96.6\%) \\
Support-assessed & Gemini 2.5 Flash & 19/29 (65.5\%) & 15/15 (100.0\%) & 4/14 (28.6\%) & 25/29 (86.2\%) \\
Support-assessed & DeepSeek-V4-Pro & 19/29 (65.5\%) & 15/15 (100.0\%) & 4/14 (28.6\%) & 23/29 (79.3\%) \\
Support-assessed & GLM-5.2 & 25/29 (86.2\%) & 14/15 (93.3\%) & 11/14 (78.6\%) & 26/29 (89.7\%) \\
\bottomrule
\end{tabular}}
\end{table*}

Exact verdict consistency ranges from 36.7\% to 93.1\%. Direct classification has the highest alert consistency in every platform and model combination. The two claim configurations use independent first-pass extractions, and support annotations do not affect the evaluated score. Differences in consistency between these configurations therefore reflect variation in claim generation and do not measure the effect of support assessment on verdicts. Exact verdict consistency is lower for malicious samples in 17 of the 18 model, platform, and decision-path combinations. The exception is direct classification with Gemini 2.5 Flash on PE, where consistency is 40.0\% for benign samples and 86.7\% for malicious samples. Most configurations are more consistent at the alert boundary than at the four-class boundary.

On PE, changing samples are closer to the benign threshold than stable samples for DeepSeek-V4-Pro and GLM-5.2, with median distances of 0.28 versus 3.40 and 0.80 versus 3.30, respectively. This relationship is absent on ELF. On both platforms, changing samples have greater variation in scores and claim counts. The fixed policy maps this variation in model-generated inputs to varying verdicts.

\subsubsection{Variation in the Accuracy Difference}
Recomputing the direct-minus-single-pass accuracy difference for each run yields medians of 16.9 points for Gemini 2.5 Flash, 17.8 for DeepSeek-V4-Pro, and 11.9 for GLM-5.2. All 30 model and run combinations favor direct classification, with a minimum difference of 8.5 points. Within-model ranges span 8.4--15.2 points. The direction persists on these 59 samples and three models, while the magnitude varies across executions. The cluster-bootstrap intervals in Table~\ref{tab:paired_stats} instead condition on the outputs of one execution and quantify variation from resampling samples.

\subsection{Replay, Rescoring, and Evidence Perturbations}

\subsubsection{Replay and Rescoring}
Two separate analyses distinguish repeated model calls from deterministic replay. For direct classification, a later execution on malicious samples provides 1,759 paired records with identical bundle hashes. Exact label agreement is 91.8\% and alert agreement is 98.0\%, with exact agreement ranging from 84.0\% for GLM-5.2 to 98.1\% for Gemma4-31B. Because execution times differ, these results combine decoding variability and possible provider-side model changes. They apply only to the malicious subset.

For single-pass processing, recomputing the evaluated policy from archived records reproduces its recorded labels in all 9,456 cases, including 2,366 from the two withdrawn models. This checks deterministic score replay and does not establish greater stability under renewed LLM calls. A stored direct label can also be retrieved unchanged. The additional capability is to recompute a verdict from the retained policy inputs under a different scoring policy.

Applying the evaluated policy to those records agrees with the deployed default on 4,682 of 9,456 labels, or 49.5\%. This disagreement reflects a policy change, not a replay failure. The archived claims and rule evidence permit that comparison without renewed model access. Direct confidence or justification could also be post-processed, but such processing would define a new decision rule rather than reconstruct the mechanism that produced the original LLM label.

\subsubsection{Evidence Perturbations}
Four fixed perturbations are applied to stored bundles. S1 removes indicator fields used by the fixed rules, S2 adds benign-looking material, S3 inserts instruction-like text into evidence strings, and S4 adds a large volume of unrelated material. S1 and S4 apply to both classes, while S2 and S3 apply only to malicious samples. A separate subset contains 182 files, with one sample per cluster and balanced classes within each architecture stratum. Including a newly executed control yields 728 bundles and 4,368 calls across Gemini 2.5 Flash, Gemma4-31B, and GLM-5.2 and the two LLM paths. The protocol was fixed before inspecting results but was not externally preregistered. Twelve responses fail to parse, or 0.3\%. The rules-only baseline is recomputed locally on the same bundles.

For the LLM paths, changes in malicious alert recall are computed on malicious samples with successfully parsed outputs in both the control and perturbed runs. The 95\% confidence intervals are the 2.5th and 97.5th percentiles of 2,000 bootstrap replicates with seed 20260827. Near-duplicate clusters are resampled with replacement, and the three models' paired outcomes are resampled together for pooled estimates. Because the 182 samples belong to distinct clusters, cluster resampling is equivalent to sample resampling.

\begin{table*}[t]
\centering
\caption{Change in malicious alert recall from the corresponding control, in percentage points with paired 95\% CIs.}
\label{tab:adv_paths}
\small
\adjustbox{max width=\textwidth}{%
\begin{tabular}{lccc}
\toprule
Condition & Rules only & Direct & Single-pass \\
\midrule
S1 Indicator hiding & $-34.1$ [$-43.9$, $-24.4$] & $-10.0$ [$-14.6$, $-5.5$] & $-29.2$ [$-36.1$, $-22.1$] \\
S2 Benign mimicry & $+0.0$ [$+0.0$, $+0.0$] & $-4.4$ [$-8.1$, $-1.5$] & $+0.4$ [$-4.1$, $+4.9$] \\
S3 Prompt injection & $+0.0$ [$+0.0$, $+0.0$] & $+3.0$ [$-0.4$, $+6.6$] & $-1.5$ [$-7.7$, $+4.5$] \\
S4 Volume dilution & $+0.0$ [$+0.0$, $+0.0$] & $+3.0$ [$-1.2$, $+7.4$] & $+3.7$ [$-2.4$, $+9.6$] \\
\bottomrule
\end{tabular}}
\end{table*}

Table~\ref{tab:adv_paths} reports changes in malicious alert recall. Under S1, malicious alert recall falls by 34.1 points for rules only, 29.2 for single-pass processing, and 10.0 for direct classification. The larger single-pass loss occurs in all four architecture strata. The rules-only baseline starts at 34.1\% recall and reaches zero, with zero ELF recall already present in the control. The experiment does not isolate claim extraction, taxonomy, and scoring as separate causes of the single-pass loss.

The rules-only outputs are unchanged under S2--S4 in these experiments. S2 reduces direct recall by 4.4 points, with its interval excluding zero, while the single-pass interval includes zero. This difference in interval significance does not itself establish a significant difference between the paths' changes. Under S4, pooled recall changes have intervals spanning zero, but benign exact-label rates fall by 5.5--11.0 points for direct classification and 4.4--9.9 for claims. Effects also vary by model, with one single-pass configuration gaining 16.7 points of recall.

Under S3, neither pooled LLM-path recall interval excludes zero. Nevertheless, single-pass processing loses alerts on 28 of 44 control-alerted malicious samples with Gemini 2.5 Flash, compared with 5 of 75 for Gemma4-31B and 0 of 64 for GLM-5.2. These conditional transition counts differ from net recall changes. The results characterize the tested bundle modifications and do not establish robustness against changes to binaries or adaptive manipulation.

\section{Discussion}
\subsection{Comparison of Decision Architectures}
Direct classification achieves higher accuracy across the six-model panel under matched evidence conditions, with positive cluster-bootstrap intervals for every paired difference in Table~\ref{tab:paired_stats}. Claim mediation also provides no consistent advantage in stability or uniformity across models. Archived Gemma3-27B results favor claims when direct classification produces substantial benign escalation, indicating that the ordering does not extend to every model examined.

The main distinction concerns the decision variables retained after inference. Claims yield higher accuracy than verdict labels in the family probe, although evidence-bundle text performs better than claims and direct justifications yield similar results. The measured benefit of the structured interface is that it retains the inputs consumed by the scoring policy. These inputs support inspection of score contributions and policy revision without another model call, including when the original model is no longer available.

\subsection{Sensitivity of the Fixed Policy}
Under 200 joint weight perturbations of up to $\pm 50\%$, with the benign threshold refitted on development data after each draw, the interval between the 5th and 95th percentiles of evaluation balanced accuracy is 3.7--8.5 percentage points wide across models. Mean perturbed performance remains within 2.5 points of the baseline for every model. These estimates characterize weight sensitivity with threshold recalibration.

Individual sweeps from $0.5\times$ to $1.5\times$ the prespecified weights produce the largest changes in panel mean balanced accuracy for benign indicators, suspicious API usage, and capability hints, at 3.0, 2.5, and 2.0 points, respectively. Each remaining category changes the panel mean by at most 1.3 points. Small changes within this range do not establish that a category is unnecessary or that its weight can be chosen arbitrarily.

Calibrated benign thresholds range from 1.75 to 6.1. Development balanced accuracy exceeds evaluation balanced accuracy by 10.6--18.3 points, whereas two-fold cross-validation within the development set reduces performance by 0.5--3.7 points. Although the latter reduction is smaller, these analyses do not provide a causal decomposition of that gap.

Selecting the threshold directly on evaluation labels increases single-pass balanced accuracy by 0--5.1 points, with a median improvement of 1.5. Direct classification still leads by 2.2--16.1 points across the six models. Thus, threshold adjustment alone does not close the gap for the retained claims and prespecified category weights. This result does not establish an upper bound for other claim representations or scoring policies.

\subsection{Implications for System Design}
The results support direct classification as the primary baseline when the required output is a sample-level label. Claim mediation serves a different operational requirement when a system needs explicit, revisable rules over retained model outputs. Category weights and thresholds can then be changed independently of claim generation. Direct confidence and justifications can also be post-processed, but that operation introduces a new decision rule rather than reconstructing the original label computation.

Under the evaluated policy, support assessment leaves verdicts unchanged for a fixed claim set. The repeated-run results characterize variability in claim generation and do not measure the effectiveness of a policy that uses support states. Dynamic evidence has mixed effects on the selected cases. The fixed scoring function supports replay, but verdicts remain dependent on the claims, categories, and severities produced by the model.

\subsection{Limitations and Threats to Validity}
The corpus is not balanced by family or collection time and does not represent operational prevalence. Family identification uses third-party annotations that were not independently verified. Difficult benign categories, including potentially unwanted programs and administrative tools, remain incompletely covered. The strong metadata baselines and unmasked identity cues also limit attribution of performance to behavioral interpretation. Product names, publishers, filenames, and embedded strings may support recognition without generalizing to unfamiliar software. Verdict accuracy is not benchmarked against external malware detectors, antivirus engines, or production analysis services.

We use a ppdeep similarity threshold of 85 for deduplication between the development and evaluation sets, compared with 70 for clustering within each corpus. The overlap audit finds no pair at 85 or above but identifies five pairs in the 70--84 range, each with the same platform and label. Checks involving benign development binaries that are no longer retained rely on archived fuzzy hashes. Residual similarity may influence threshold calibration. The oracle analysis preserves the observed ranking for the stored scores but does not establish that the development split is independent or free of overlap.

Matching evidence and models does not isolate representation format from the other differences between the paths. Prompts, response schemas, output budgets, parsing, and verdict assignment differ, and only the claim path uses a development-calibrated threshold. The paired analysis excludes a sample from both paths if either lacks a usable output, whereas full-corpus metrics retain errors. The findings therefore concern the implemented decision architectures under their reported configurations.

Collected evidence reflects the selected tools and observation conditions. Static analysis, decompilation, and runtime collection may each omit relevant behavior, and the bundles were not validated against independent reverse engineering. The study also does not measure semantic claim precision, completeness, expert agreement, or effects on analyst performance. The demonstrated linkage to score computation does not establish the correctness or practical usefulness of individual claims.

The perturbation study evaluates four fixed modifications to stored bundles. It does not establish that these modifications can be realized through binary changes or characterize adaptive manipulation of the collector and decision paths. The observed differences in sensitivity therefore apply to the tested conditions and do not identify the causal contribution of extraction, taxonomy, or scoring.

Provider withdrawal prevents renewed claim extraction for two models, and floating model tags limit exact re-execution for the remaining panel. Archived claims and evidence permit policy replay and rescoring, but they do not recover the original model service or guarantee identical outputs from future calls.

\section{Related Work}
LLMs have been applied to several stages of malware analysis. Fujii and Yamagishi examined their feasibility for static malware analysis, while Patsakis et al.\ evaluated deobfuscation of malicious code from real campaigns~\cite{fujii2024feasibility,patsakis2024assessing}. MaLAware generates behavioral summaries from Cuckoo Sandbox reports~\cite{saha2025malaware}. For Android, Walton et al.\ combined semantic summarization with malware categorization, and He et al.\ introduced Cama to benchmark malicious function identification and malware purpose summarization~\cite{walton2024exploring,he2025benchmarking}. These Android studies concern application code and platform-specific evidence, whereas the present study evaluates evidence collected from native PE and ELF binaries.

Research on binary understanding addresses related capabilities at the code level. LLM4Decompile evaluates decompilation through the executability of reconstructed code~\cite{tan2024llm4decompile}. Shang et al.\ studied function-name recovery and binary code summarization, and subsequently introduced BinMetric to assess a broader set of binary-analysis tasks~\cite{shang2024far,shang2025binmetric}. Xuan et al.\ proposed TTPDetect, an LLM agent that identifies adversary tactics and techniques in decompiled functions~\cite{xuan2026identifying}. TTPDetect evaluates the recognition and localization of behaviors within code, whereas MARS examines how behavioral claims contribute to a security verdict for an entire sample.

Concept bottleneck models provide a related formulation in which predicted, human-specified concepts are the sole inputs to a downstream classifier~\cite{koh2020concept}. The evaluated claim-based path in MARS similarly restricts verdict assignment to extracted claims. However, these claims are produced by a general LLM without concept-level supervision, and the verdict policy uses a prespecified scoring rule with thresholds calibrated on development data.

ElZemity et al.\ examine a different architectural choice by comparing orchestration strategies for multiple models, including evidence pipelines, adversarial debate, and consultation between general and specialized models~\cite{elzemity2026small}. Their evaluation measures accuracy on multiple-choice questions about malware detonation reports. Their focus is coordination among models, while ours is the placement of verdict assignment within a pipeline.

The perturbation study relates to both malware evasion and prompt injection. Demetrio et al.\ survey and evaluate functionality-preserving changes to Windows binaries that evade static machine-learning detectors~\cite{demetrio2021adversarial}. Perez and Ribeiro study goal hijacking and prompt leaking, while Greshake et al.\ demonstrate indirect prompt injection through data consumed by LLM applications~\cite{perez2022ignore,greshake2023not}. This threat is relevant to malware analysis because strings and metadata supplied to the model can originate from attacker-controlled artifacts. Our study applies predefined perturbations to stored evidence bundles and compares the resulting verdict changes across decision architectures. Because the binaries remain unchanged, these experiments characterize sensitivity to altered evidence rather than establish executable malware evasion.

The present study therefore compares decision architectures for malware triage using a shared evidence collector and corpus. Direct classification and claim-based classification receive the same serialized evidence for each model, with deterministic rules providing a reference baseline. The contribution is an empirical assessment of the consequences of assigning verdicts through intermediate behavioral claims, including classification performance, stability, and sensitivity to evidence perturbations.

\section{Conclusion}
This paper presented MARS, a malware triage framework for PE and ELF binaries, and evaluated the consequences of separating LLM evidence interpretation from verdict assignment. Using identical serialized evidence bundles, we compared direct classification with single-pass claim extraction followed by a fixed scoring policy across six models and 1,195 binaries, alongside a deterministic rules-only baseline. Under the evaluated claim taxonomy and additive policy, direct classification achieved higher overall accuracy for all six models and higher malicious alert recall in ten of twelve platform and model combinations. Claim mediation did not consistently reduce variation across models. The claim path nevertheless makes verdict computation inspectable and revisable. Because the retained claim categories and severities are the policy inputs, their contributions can be examined and verdicts can be recomputed without another model call. Exact replay of archived verdicts remained possible for two models withdrawn by the provider, and the same records supported rescoring under a revised policy. These properties did not ensure stability or robustness. Repeated execution showed that a deterministic policy still produces varying verdicts when its model-generated inputs change. In the predefined evidence perturbation study, removing selected indicator fields caused a larger loss of malicious alert recall for the claim path than for direct classification. These findings concern the tested changes to stored evidence and do not establish robustness against modifications to executable binaries. Under the evaluated configurations, the results favor direct classification when only a sample-level verdict is required. Claim mediation provides explicit policy control, whose practical value must be weighed against the observed costs in accuracy and stability. Future work will extend the comparison to evidence from production analysis platforms, evaluate both paths against adaptive adversaries with knowledge of the decision policy and against modifications to binaries before evidence collection, and measure whether retained claims improve analyst decisions.

\bibliographystyle{elsarticle-num}
\bibliography{bib}

\end{document}